\documentclass{amsart}

\usepackage{amssymb}    

\theoremstyle{plain}
\newtheorem{thm}{Theorem}[section]

\theoremstyle{definition}
\newtheorem{ex}[thm]{Example}

\newtheorem{defn-re}[thm]{Definition and Remark}

\begin{document}

\title{Periodicity and Growth in a Lattice Gas With Dynamical Geometry}

\author{Karin Baur}
\address{Department of Mathematics, University of California, San Diego,
La Jolla, CA 92093 USA}
\thanks{Karin Baur was supported by Freie Akademische Stiftung and 
by DARPA contract \#AFRL F49620-02-C-0010. Address after June 30, 2005: 
Department of Mathematics, University of Leicester, University Road,
Leicester LE1 7RH, UK. Email kubl@mcs.le.ac.uk}
\email[Baur]{kbaur@math.ucsd.edu}

\author{Jeffrey M. Rabin}
\email[Rabin]{jrabin@ucsd.edu}

\author{David A. Meyer}
\thanks{David Meyer was supported by AFOSR grants F49620-01-1-0494 and
FA9550-04-1-0466}
\email[Meyer]{dmeyer@ucsd.edu}

\date{June 30, 2005}

\maketitle
%
\section*{Abstract}
%

We study a one-dimensional lattice gas ``dynamical geometry model" in which
local reversible interactions of counter-rotating groups of particles
on a ring can create or destroy lattice sites.
We exhibit many periodic orbits and show that all other solutions have
asymptotically growing lattice length in both directions of time.
We explain why the length grows as $\sqrt{t}$ in all cases examined.
We completely solve the dynamics for small numbers of particles with
arbitrary initial conditions.

\vskip 40pt


\section{Introduction}


Lattice models are ubiquitous in physics, whether as regularizations for
continuum theories (quantum field theory, quantum gravity), scaffolding
for numerical methods (classical field theories, continuum mechanics), or
because the lattice is physically real (condensed matter physics).
In virtually all applications, however, the lattice structure and size are
fixed, or at least not dynamical.
(In numerical computation the lattice may be refined to maintain precision, 
but the evolution of the lattice is not part of the physical dynamics
under study.)
Among the exceptions known to us are the causal dynamical triangulation
approach to quantum gravity \cite{ajl05} and the variable-length lattice 
models of recent relevance to the AdS/CFT correspondence \cite{bcv05}.

In \cite{hm98}, Hasslacher and Meyer constructed a lattice gas model with
dynamical geometry and a reversible evolution rule.
It can be viewed as a toy model for general relativity in that the geometry 
(length) of the one-dimensional lattice changes in response to the motion 
(scattering) of the matter particles on it.
The model is classical, but its quantization should present no problems.
In this paper, we extend the previous analyses of the classical dynamics.
The central issue is the long-time behavior of the lattice length.
We will completely solve the dynamics for systems of a few particles, and
explain the typical $\sqrt{t}$ growth of the length which has been
seen previously in simulations.

The model consists of a one-dimensional lattice of $L$ sites with periodic
boundary conditions (a ring), where $L$ may change with time.
The initial state contains $N_R$ right-moving particles and $N_L$
left-moving ones, which may be placed arbitrarily on the sites subject to an
exclusion principle: two particles moving in the same direction may not
occupy the same site.
The numbers of left- and right-movers are each conserved during evolution.
Time proceeds in discrete steps.
At each time step, the particles first {\it advect}: each particle moves one
site in its own direction of motion.
Then the particles {\it scatter} according to the rules (see Figure 1):

\begin{figure}
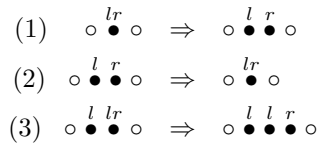

\begin{eqnarray*}
(1) \;\;\;\;\; \circ \stackrel{lr}{\bullet} \circ & \Rightarrow & \circ \stackrel{l}{\bullet}
\;\stackrel{r}{\bullet} \circ \\
(2) \;\;\; \circ \stackrel{l}{\bullet}\; \stackrel{r}{\bullet} \circ & \Rightarrow &
\circ \stackrel{lr}{\bullet} \circ \\
(3)\;\;\;  \circ \stackrel{l}{\bullet}\; \stackrel{lr}{\bullet} \circ & \Rightarrow & 
\circ \stackrel{l}{\bullet}\; \stackrel{l}{\bullet}\; \stackrel{r}{\bullet}
\circ \\
\end{eqnarray*}
\caption{The three scattering rules. Filled-in sites are occupied. {\it l}
denotes a left-mover, {\it r} a right-mover.}
\end{figure}

(1) If a right- and left-mover occupy the same site, this site is replaced
by two sites, with the right-mover on the rightmost site and the left-mover
on the leftmost site.

(2) If two adjacent sites are occupied, with a right-mover
on the rightmost site and a left-mover on the leftmost, these sites are
replaced by a single site occupied by both particles.

(3) Rules (1) and (2) both apply when two adjacent
sites are occupied by three particles (or by four), 
with the singleton moving away from the doubly-occupied site.
Since the application of rule (2) would violate the Exclusion Principle, 
in such cases {\it only} rule (1) is applied.

These rules define a reversible dynamics in that every state has not only a 
unique successor but also a unique predecessor.

Rule (3), the {\it exclusion rule}, is the major complication in analyzing the 
dynamics, as will be seen in section 2 below.
The situation it describes can arise following advection when a single particle
moving in one direction approaches a pair of particles on adjacent sites
moving in the other.
Such a pair of particles moving in the same direction on adjacent sites will
be called simply a {\it pair}, and we will see that such ``bound states" 
can play the role of quasiparticles in the system.

After initial explorations in \cite{hm98}, numerical simulations of the 
evolution of initial states rather densely populated with particles 
(25 particles on a 50 site lattice) were carried out in \cite{lbm04}.
Despite the reversibility of the dynamics, most initial states result in
growth of the lattice size, empirically as $L(t) \sim \sqrt{t}$ at late times.
On small lattices, some ``rogue states" were also found with $L(t)$ periodic,
but the proportion of these dropped off rapidly with initial lattice size.
Two versions of mean-field theory were proposed to explain the observed
growth rate, one of which predicted $\sqrt{t}$ growth as observed while
the other predicted $t^{1/3}$.
In this paper we will identify many periodic solutions on lattices of
arbitrary size, and also propose an alternative explanation for the
typical $\sqrt{t}$ growth.

The reversibility of the dynamics leads to the following simple but
fundamental

{\bf Evolution Theorem:} {\it Every solution of the model is either periodic,
or grows without bound in both directions of time.}

{\it Proof:} Consider a solution for which the lattice length remains bounded
in one direction of time, say $t \rightarrow \infty$. 
Since there are only finitely many distinct states of this system on a lattice
of given size, the evolution must eventually return to some previous state.
The evolution is then periodic from this time on.
By reversibility and uniqueness, it is periodic in backward time also.

The organization of this paper is as follows.
In section 2 we give a general analysis of the evolution of initial states,
assuming that no pairs are present or form later.
We point out the crucial role of parity in the problem (first noted
in \cite{lbm04}) and establish the existence of many periodic solutions.
We explain why growing solutions of this kind must grow as $\sqrt{t}$.
In section 3 we exhaustively analyze the evolution of all initial states
containing at most four particles, possibly including pairs.
We see cases in which permanent pairs form and behave as quasiparticles or
bound states in the system.
The major obstacle to a complete solution of this model is the lack of a
general framework for describing these quasiparticles and their effects.
Section 4 contains conclusions and open problems.

We would like to acknowledge discussions at the early stages of this work with
the authors of \cite{nlb05}, who have independently obtained similar results.

%
\section{General Analysis of States Without Nearest-Neighbor Pairs}
%

Although simulations of the time evolution of ``random" initial states
look quite complicated, there are several important general principles
governing the dynamics which can be formulated.
First, the translational (rotational) symmetry of the lattice allows
the dynamics to be viewed in various reference frames.
We have thus far used a frame fixed with respect to the lattice, but we
can transform to the rest frame of either the right- or the left-moving
particles.
The rest frame of the left-movers, for example, is defined as follows.
At each advection step, the left-movers move one site to the left.
We can follow this advection step with a symmetry (gauge) transformation
which rotates every particle one step to the right, thus undoing
the advection for the left-movers.
This is followed by the scattering step as usual.
In this frame, or gauge, the left-movers do not advect, while
the right-movers advect {\it two} sites per time step.

The use of the rest frame for one group of particles makes it clear that
parity (mod 2) plays a crucial role in the dynamics.
The size of the gap between two left-movers, or two
right-movers, is trivially preserved by advection, but so is the parity of
the (shrinking) interval between a left- and a right-mover.
If this interval contains no other particles,
its parity determines the character of the eventual interaction between
these particles, assuming neither belongs to a nearest-neighbor pair:
a site will be created (resp., destroyed) if this parity is even (resp., odd).
In turn, either type of interaction will reverse the parity of a gap containing
the newly created or destroyed site.
Thus, we can reduce the dynamics mod 2, and study the evolution of the gap
parities according to these simple rules.
The presence of pairs can invalidate the analysis by preventing destruction
events which would otherwise occur.
The corresponding parity reversals then do not take place.

Thus, our plan for analyzing the dynamics is to assume first that no pairs
are present, and no pairs form during evolution.
Analysis via parity then leads to very simple and general results, which are
valid if they are self-consistent, that is, if they do not lead to the
formation of pairs.
We then consider the excluded cases in which pairs are present initially
or form later.
The pairs can violate the parity ``selection rules", causing transitions
between the types of behavior observed when those rules hold.
At present, we can only analyze the effects of pairs in a laborious
case-by-case manner with small numbers of particles.
An important task for the future is to develop appropriate concepts for
a general analysis of the effects of pairs, which might be thought of as
quasiparticles in the system.

To introduce the technique of parity analysis, consider the case of one
right-mover versus a string of $N \equiv N_L$ left-movers, without pairs.
We work in the rest frame of the left-movers, so the right-moving particle
advects by two sites per time step, preserving parities.
We label the left-movers from 1 to $N$ in the order that the right-mover will
encounter them, and describe the system by the gaps
$g_{12}, g_{23}, \ldots, g_{N1}$ between successive left-movers as well as
the parity $p$ of the separation between the right-mover and the first
left-mover.
(The term ``gap" always refers to the distance between consecutive left-movers,
or consecutive right-movers; we use the term ``separation" for the shrinking
distance between a left-mover and right-mover before they scatter.)
This data is preserved by advection, and changes as follows under scattering:
$$
[p; g_{12}, g_{23}, \ldots, g_{N1}] \rightarrow
[|p+g_{12}|; g_{23}, \ldots, g_{N1}, g_{12}+(-1)^p],
$$
where $|x|$ is the parity of $x$.
The gaps are always listed in the order that the right-mover will encounter
them.

Let the right-mover make a complete circuit, passing all $N$ left-movers and
returning to face the first again.
(Note: this complete circuit in the rest frame is only a half-circuit in
the lattice-fixed frame.
In that frame, the right-mover passes each left-mover twice as every particle
makes a complete circuit of the lattice.)
Each gap parity $g_{i,i+1}$ has been reversed, and $p$ has changed by
the sum of the original gap parities, which is the parity of the original
lattice length $L$.
(It may be surprising that the parity of $L$ prior to the complete circuit
determines whether the right-mover returns to its original position
relative to the first left-mover or is displaced from it by one site.
The point is that sites which may be created or destroyed during this circuit
do not increase or decrease the distance the right-mover must advect to
finish the circuit.
This is because the right-mover is carried forward or backward with the
newly created or destroyed site as part of the scattering step.
Changes in $L$ during this circuit take effect at the next circuit.)
The parity of the lattice length itself has changed by $|N|$ due to the $N$
scattering events.
Now let the right-mover make another circuit.
This restores all the original gap parities $g_{i,i+1}$, but not necessarily
$p$, which differs from its original value by $|2L+N|=|N|$.

Supposing first that $N$ is even, the system is parity-periodic with period
two circuits, that is, $2N$ interactions.
This implies that further circuits will repeat the same pattern of creation
and destruction interactions as the first two circuits.
However, the actual gap lengths are not periodic, because the interactions
of the second circuit do not undo the effects of the first.
If, for example, $L$ is odd, it is easy to see that $g_{12}$ is unchanged
after two circuits while $g_{23}$ has changed by $\pm 2$.
Indeed, regardless of $|L|$, alternate gaps have changed by
$0, \pm 2, 0, \pm 2, \ldots$ after two circuits.
The system either grows forever (if all signs are $+$) or a pair
eventually forms and invalidates the analysis.

Suppose next that $N$ is odd.
After two circuits $p$ has reversed parity while all other gaps have their
original parities.
It follows that the interactions of circuits three and four will exactly
undo the effects of circuits one and two respectively, and the evolution will
be truly periodic with period four circuits, or $4N$ interactions.
During these four circuits, each gap can change by at most $\pm 2$ from its
initial size.
Therefore, $g_{i,i+1} > 3$ is sufficient to prevent the formation of
pairs and render this behavior self-consistent.
We have thus established the existence of a large class of periodic solutions.

Note that there are $2^{N+1}$ possible parity states, given by the
parities of the $N$ gaps and of $p$.
Since this number is finite, periodic evolution of the parities is
inevitable.
Since the period, $2N$ or $4N$, only grows linearly with $N$,
the $2^{N+1}$ states clearly belong to a large number of disjoint parity
orbits when $N$ is large.
Some examples with $N$ small are given in the next section.

With a mild additional assumption this analysis can be extended to the
general case of $N_R$ right-movers versus $N_L$ left-movers, of course
assuming the absence of pairs.
Provided that no right-mover is very near a left-mover in the initial state,
we can again consider a complete circuit in which every right-mover passes
every left-mover exactly once.
Each right-mover interacts $N_L$ times, making $N_R N_L$
interactions in all.
In the rest frame of the left-movers, it is clear that each gap
between consecutive left-movers changes parity by $|N_R|$ during one
circuit of the right-movers.
Similarly, each gap between consecutive right-movers changes parity by
$|N_L|$.
There is also a parity change in the separation between a chosen left-mover
and a chosen right-mover after one circuit.
It suffices to compute this for one such choice, because the others are
then determined by the known gap changes.
However, this ``offset parity" depends on the exact configuration
of the particles around the ring.
We will always work in the rest frame of the left-movers, and compute the
offset between a chosen right-mover and the first left-mover it will
encounter.

For example, suppose that initially one arc of the ring contains all the 
left-movers and no right-movers, and another contains all the right-movers.
Consider the leading right-mover.
As it passes the left-movers, it is carried along with the creation and
destruction events, and its offset parity after one cycle is simply $|L|$.
Contrast this with an initial configuration in which $N_L=N_R=N$ and
the left- and right-movers alternate around the ring.
Choose a ``leading" right-mover arbitrarily.
Number the left-movers in the order that this right-mover will encounter
them.
It passes the first and traverses the gap $g_{12}$, but the next gap has
already been changed to $g_{23} \pm 1$ by the interaction with
the right-mover ahead of the chosen one.
Similarly the next gap will be $g_{34} \pm 1 \pm 1$ (independent signs) 
when the chosen particle gets there, and the offset has the parity of
$|L + 1 + 2 + \cdots + (N-1)| = |L + {1 \over 2} N(N-1)|$.
In general, the offset is $L$ plus the number of interactions with left-movers
which occur before the chosen right-mover reaches them.
The latter contribution cancels out (mod 2) after two circuits.

Now we analyze the various parity combinations in detail.
Suppose first that both $N_R$ and $N_L$ are even.
Then the parity of every gap is unchanged after one circuit.
If the offset parity for some particle is also even, then it is even
for every particle, and the pattern of
interactions at every successive circuit is identical.
The system either
grows indefinitely or eventually forms a pair and invalidates the analysis.
If, however, the offset for some (hence every) particle was odd,
then the interactions of the second circuit undo the effects of the first,
and the system is truly periodic with period two circuits or
$2 N_R N_L$ interactions.
No gap changes by more than $\max ( N_R, N_L )$ during the evolution.

Next suppose that $N_R$ and $N_L$ are both odd.
Now every gap changes parity in a circuit,
as does the lattice length.
After two circuits, the gap parities have their
original values, but the offset is odd.
As in the case of $1$ versus odd $N$ above, circuits three and four undo the
effects of circuits one and two, and we get truly periodic solutions with
period four circuits or $4 N_R N_L$ interactions.

Finally, suppose $N_R$ is odd and $N_L$ is even (the opposite case being
the same by symmetry).
After one cycle the left-left gap parities are reversed, while the
parities of the right-right gaps and the lattice length are unchanged.
As in the case of 1 versus even $N$ above, the interactions of the next
cycle cancel those of the first for every other left-left gap, but
augment those of the first for the remaining left-left gaps.
The result is either net growth or pair formation.

To summarize, in the absence of pairs periodic solutions are quite generic
in the cases $N_R, N_L$ both odd, and both even with offset parity odd.

Now consider any solution which grows indefinitely.
Because the gap and offset parities repeat after a certain number of circuits,
and determine the pattern of interactions and thus the net
number of sites created, the growth is characterized by some average number
of sites created per circuit.
Let $k$ be this number, and $L(t)$ be the lattice length at time $t$.
Since a circuit takes $L/2$ time steps,
in the limit of continuous time we have the differential equation
$$ {dL \over dt} = {2k \over L},$$
with asymptotic solution $L \sim 2\sqrt{ k t}$.
The $\sqrt{t}$ growth observed in simulations thus simply reflects
constant average growth per circuit.

It is tempting to claim that this reasoning is completely general, applying
even if nearest-neighbor pairs form.
The argument would be that the pattern of interactions is still determined by  
a finite set of data, namely the parities of all gaps and an offset, 
and a list of which particles are paired.
As this finite set of data changes, it must eventually return to a former
state, from which point the pattern of interactions will be periodic.
There will be a net number of sites created per period, leading again to the
$\sqrt{t}$ growth, on a sufficiently long time scale.
However, this finite set of data is in fact insufficient to determine the
sequence of interactions, because one needs the actual gap sizes, not just
their parities, to predict when a new pair will form.
Thus, at present we cannot prove that every nonperiodic solution grows
according to the $\sqrt{t}$ law.

\vskip 18pt

%
\section{Few-Particle Systems}
%
%

\begin{figure}
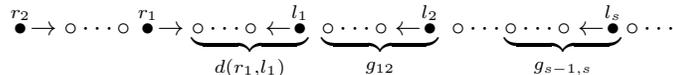
\[
\quad\stackrel{r_2}{\bullet}\to
\circ
\dots\circ\stackrel{r_1}{\bullet}\to
\underbrace{\circ\dots\circ
\leftarrow \stackrel{l_1}{\bullet}}_{d(r_1,l_1)}
\ \underbrace{\circ\dots\circ\leftarrow
\stackrel{l_2}{\bullet}}_{g_{12}}\ \circ\cdots
\underbrace{\circ\cdots\circ
\leftarrow\stackrel{l_s}{\bullet}}_{g_{s-1,s}}
\circ\cdots
\]
\caption{Notation for interparticle gaps and separations.}
\end{figure}

In this section we will completely solve the dynamics for all initial states
containing at most four particles.
By symmetry we may assume at most two are right-movers; for the moment
we consider a single right-mover with up to three left-movers.
We describe the states of the system at time $t$
by the parity of the separation between the right-mover and the
left-mover it will encounter next,
and by the gaps $g_{i,i+1}$ between left-movers $i$ and $i+1$
(where the indices $i,i+1$ are taken modulo $N_L$). 
After every interaction we give the new state.

\subsection{One Against $N_L$}

\begin{ex}
{\bf One against one}

For completeness and to establish notation we begin with the trivial case
$N_R=N_L=1$.
We keep track of the parity $p$ of the separation
$d(r,l)$ and the parity of the gap $g_{11}$,
in other words of the length of the lattice.
The state is described by $[p,|L|]$.
To the right we give the value of the growth $\Delta L$ following each 
interaction.

\[
\begin{array}{cc}
\left[1;1\right] & \\
& \downarrow -1 \\
\left[0;0\right] & \\
& \downarrow +1 \\
\left[0;1\right] & \\
& \downarrow +1 \\
\left[1;0\right] & \\
& \downarrow -1 \\
\left[1;1\right] &
\end{array}
\]

Any state in this cycle can be viewed as the ``initial" state;
all possible parities for one against one belong to the same orbit.
This orbit has net growth zero, and is truly periodic.
\end{ex}

\begin{ex} {\bf One against two}

The first interesting case is one right-mover against two left-movers.
Here, the exclusion rule can apply (nearest-neighbor pairs suppressing the 
destruction of sites). 

We describe the possible states as follows:
$\left[p;g_{i,i+1},g_{i+1,i+2};|L|\right]$,
where $p$ is the parity of the separation
$d(r,l_i)$ between $r$ and the closest left
mover $l_i$ ($i=1,2$), and $|L|$ is the parity of the lattice length. 

With this data we can keep track of the position
of $r$: we can always see which particles interact
next.
Note that $|L|=|g_{12}+g_{21}|$.
For convenience we display parity by using bold letters ${\bf g_{i,i+1}}$ 
if and only if the gap $g_{i,i+1}$ is odd.

There are eight possible combinations of the
parities of $d(r,l_i)$, $g_{i,i+1}$, $g_{i+1,i+2}$.
We start with the two states (1) and (i), 
assuming at first that no pairs occur (see below).
\[
\begin{array}{clr||clr}
(1)&\left[1;{\bf g_{12}}, g_{21};1\right] &
&(i)&\left[1;g_{12},g_{21};0\right] &  \\
& & \downarrow -1 && & \downarrow -1 \\
(2)&\left[0; g_{21}, g_{12}-1;0\right] &
&(ii)&\left[1;g_{21},{\bf g_{12}-1};1\right]& \\
& & \downarrow +1 && & \downarrow -1 \\
(3)&\left[0; g_{12}-1,{\bf g_{21}+1};1\right] &
&(iii)&\left[1;{\bf g_{12}-1},{\bf g_{21}-1};0\right] & \\
& & \downarrow +1 && & \downarrow -1 \\
(4)&\left[0;{\bf g_{21}+1},{\bf g_{12}};0\right] &
&(iv)&\left[0;{\bf g_{21}-1},g_{12}-2;1\right] &\\
& & \downarrow +1 && & \downarrow +1 \\
(5) &\left[1;{\bf g_{12}}, g_{21}+2;1\right] &
&(v)&\left[1;g_{12}-2,g_{21};0\right]&
\end{array}
\]

In this chart, the quantities $g_{12}$ and $g_{21}$ denote the {\it initial}
values of these gaps; current values are indicated by position within 
the brackets.
For example, line (3) indicates that the right-mover is about to encounter
$l_1$, the gap from $l_1$ rightward to $l_2$ is currently $g_{12}-1$,
and the gap from $l_2$ rightward to $l_1$ is $g_{21}+1$.
Observe that the eight states
(1)-(4) and (i)-(iv) cover all the parity configurations.
Thus if no pairs occur, these eight states
describe all possible behaviors of one particle against two.

Consider the first example, states (1) - (5).
Note that the parities of state (5) are
a repetition of the parities of state (1).
There is one gap that has grown by two,
$g_{21}\mapsto g_{21}+2$,
while the other gap remained constant. 
So this is a parity-periodic growing orbit as long as no pairs appear. 
The only places where a pair can form are in (1), (5), (9), and so on.
The restriction $g_{12}\ge 3$
prevents the formation of pairs.
In particular, we have a growing orbit if and only if $g_{12}\ge 3$.
The growth rate is as $\sqrt{t}$ as discussed previously.

The second example describes a shrinking system.
After four interactions we get back to the same
configuration of parities but with a shorter lattice. 
Two sites have been eliminated.
It is clear that eventually pairs
will form and alter the evolution in the states (iii), (vii), (xi), etc.
where we have an odd distance of $r$ to $l_1$.
Although we will consider the effects of these pairs below, the eventual
fate of this system can be determined by the following time-reversal
argument.
Running time backward from the initial state, we would obviously see
this system grow, with no pair formation.
By the Evolution Theorem, this system must eventually grow in the future
as well.

To complete the analysis of one right-mover against two left-movers, 
we now describe what happens if pairs form. 
In that case $g_{12}=1$, and we can assume that the distance $d(r,l_1)$ 
is odd, since otherwise the existence of the pair does not affect the 
evolution.
Since we have the choice of the parity of $g_{21}$,
there are two types of states with a pair:

\[
\begin{array}{lr||lr}
(A) &  & (B) & \\
 & & & \\
\left[1;{\bf g_{12}=1},g_{21};1\right]  & &
\left[1;{\bf g_{12}=1},{\bf g_{21}};0\right] & \\
  & \downarrow {\bf +1} & & \downarrow {\bf +1} \\
\left[0;{\bf g_{12}=1},{\bf g_{21}+1};0\right] &
&\left[1;{\bf g_{12}=1},g_{21}+1;1\right] &
\end{array}
\]

In the above evolution, the boldface values of $\Delta L = {\bf +1}$
indicate steps at which the evolution was altered by the presence of the
pair.
Note that (A) has evolved into step (4).
Since in (A) we started with $g_{21}\ge 2$,
we get $g_{21}+1\ge 3$ and remain in the growing orbit.
Case (B) has evolved into (A) after one step, so both
(A) and (B) evolve to the growing orbit.
The pair breaks, and the gap between these neighbors subsequently grows.
\end{ex}

Now let $r$ face a sequence of left-movers, $l_1,\dots, l_{N_L}$, 
the initial state being $[p;g_{12},g_{23},\ldots,g_{N_L1}]$.
If particles 1 and 2 do not form a pair, then the state following the first
interaction is as given in the previous section, namely
$$[|p+g_{12}|;g_{23},\ldots,g_{N_L1},g_{12}+(-1)^p].$$
If particles 1 and 2 do form a pair, this changes the outcome iff $p=1$.
In that case, $r$ does not interact with $l_1$, and following the interaction
with $l_2$ the state will be 
$$[|g_{23}|;g_{34},\ldots,g_{N_L1},g_{12},g_{23}+1].$$

The interaction between $r$ and a single pair can be conceptualized 
as follows.
If the gap $d(r,l_1)$ to the leading member of the pair is odd, the
right-mover lands on the trailing member of the pair.
The pair remains intact and a site is created behind it, so that
$\Delta L =1$.
In this case the pair behaves as a unit like a single left-mover.
If the gap $d(r,l_1)$ is even, the right-mover lands on the leading member
of the pair.
A site is created between the left-movers, breaking the pair, and another
site is destroyed behind the pair, resulting in $\Delta L=0$.
More globally, under the right conditions a pair may persist indefinitely,
behaving like a quasiparticle in the system, as we will see later.
Or a pair might break and re-form repeatedly.
Notice that, in contrast, three left-movers on adjacent sites cannot form a 
stable ``triple"; it will necessarily be broken by the next interaction with a
right-mover.

\begin{ex} {\bf One against three}

Let us turn to the case of one right-mover against three left-movers.
\begin{figure}
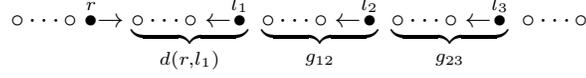

\[
\quad\circ\dots\circ\stackrel{r}{\bullet}\to
\underbrace{\circ\dots\circ
\leftarrow \stackrel{l_1}{\bullet}}_{d(r,l_1)}\
\underbrace{\circ\dots\circ\leftarrow
\stackrel{l_2}{\bullet}}_{g_{12}}\
\underbrace{\circ\dots\circ\leftarrow
\stackrel{l_3}{\bullet}}_{g_{23}}
\ \circ\dots\circ
\]
\caption{Notation for one right-mover vs. three left-movers.}
\end{figure}
Here we describe a state by the parity $p$ of $d(r,l_1)$
(assuming $r$ faces $l_1$ next), by $g_{12}$, $g_{23}$, $g_{31}$
and (for convenience, although it is redundant) 
the parity $|L|$ of the length of the lattice.
There are sixteen possible parity configurations for the
gaps and separation $p$.

We obtain two periodic orbits of length $12$
under the assumption that no pairs form.
We give them here:

\[
\begin{array}{rlc|rlc}
(1) & [1;g_{12},g_{23},g_{31};0]
&&(a) & [1;g_{12},{\bf g_{23}},g_{31};1]  \\
&& -1 \downarrow &&& -1\downarrow   \\
(2) & [1;g_{23},g_{31},{\bf g_{12}-1};1] &
&(b) & [1;{\bf g_{23}},g_{31},{\bf g_{12}-1};0] & \\
&&-1 \downarrow && & -1 \downarrow  \\
(3) & [1;g_{31},{\bf g_{12}-1},{\bf g_{23}-1};0] &
&(c) & [0;g_{31},{\bf g_{12}-1},g_{23}-1;1] & \\
&&-1 \downarrow  && & +1 \downarrow  \\
(4) & [1;{\bf g_{12}-1},{\bf g_{23}-1},{\bf g_{31}-1};1] &
&(d) & [0;{\bf g_{12}-1},g_{23}-1,{\bf g_{31}+1};0] & \\
&&-1 \downarrow && & +1 \downarrow  \\
(5) & [0;{\bf g_{23}-1},{\bf g_{31}-1},g_{12}-2;0] &
&(e) & [1;g_{23}-1,{\bf g_{31}+1},g_{12};1] & \\
&&+1 \downarrow  && & -1 \downarrow  \\
(6) & [1;{\bf g_{31}-1},g_{12}-2,g_{23};1] &
&(f) & [1;{\bf g_{31}+1},g_{12},{\bf g_{23}-2};0] & \\
&&-1 \downarrow && & -1 \downarrow  \\
(7) & [0;g_{12}-2,g_{23},g_{31}-2;0] &
&(g) & [0;g_{12},{\bf g_{23}-2},g_{31};1] & \\
&&+1\downarrow & & &+1 \downarrow  \\
(8) & [0;g_{23},g_{31}-2,{\bf g_{12}-1};1] &
&(h) & [0;{\bf g_{23}-2},g_{31},{\bf g_{12}+1};0] & \\
&&+1 \downarrow  && &+1 \downarrow  \\
(9) & [0;g_{31}-2,{\bf g_{12}-1},{\bf g_{23}+1};0] &
&(i) & [1;g_{31},{\bf g_{12}+1},g_{23}-1;1] & \\
&&+1 \downarrow  & && -1 \downarrow  \\
(10) & [0;{\bf g_{12}-1},{\bf g_{23}+1},{\bf g_{31}-1};1]\quad &
& (j)& [1;{\bf g_{12}+1},g_{23}-1,{\bf g_{31}-1};0] & \\
&&+1 \downarrow && & -1 \downarrow  \\
(11) & [1;{\bf g_{23}+1},{\bf g_{31}-1},g_{12};0] &
& (k) & [0;g_{23}-1,{\bf g_{31}-1},g_{12};1] & \\
&&-1 \downarrow  && & +1 \downarrow  \\
(12) & [0;{\bf g_{31}-1},g_{12},g_{23};1] &
& (l) & [0;{\bf g_{31}-1},g_{12},{\bf g_{23}};0] & \\
&&+1 \downarrow && & +1 \downarrow  \\
(1) & [1;g_{12},g_{23},g_{31};0] &&
(a) & [1;g_{12},{\bf g_{23}},g_{31};1]
\end{array}
\]

The first system is free of pairs if and only if $g_{12}\ge 4$,
$g_{23}\ge 2$ and $g_{31}\ge 4$:
the only cases where the exclusion rule can apply are in
steps (4) if $g_{12}-1=1$, in (6) if
$g_{31}-1=1$ and in (11) if $g_{23}+1=1$.
(Of course, this would literally imply $g_{23}=0$, which is not possible 
by the Exclusion Principle.
What is meant is that an initial state with the parities of line (11) would
contain a pair if the entry $g_{23}+1$ were $1$ instead.)

The second system is free of pairs if and only if
$g_{12}\ge 2$, $g_{23}\ge 3$, $g_{31}\ge 2$.
The only cases where the exclusion rule can apply are
in steps (b) if $g_{23}=1$, in (f) if
$g_{31}+1=1$ and in (j) if
$g_{12}+1=1$.

Note that all sixteen parity configurations
appear in the two orbits.
The first orbit covers $12$ parity configurations.
In the second orbit, the parity configurations repeat after $4$ steps, 
but the actual gap sizes have period $12$.

What is left is to understand the
cases where pairs form (and thus the exclusion rule applies).
In other words we have to study the systems with
$d(r,l_1)$ odd, $g_{12}=1$ and all possible
parities of $g_{23}$, $g_{31}$. 
We give them labels as follows:
\[
\begin{array}{cc}
  \mbox{parities of}\ (g_{23},g_{31}) &
  \mbox{label}\\
\hline
\phantom{} [0,0] & (A) \\
\phantom{} [0,1] & (B) \\
\phantom{} [1,0] & (C) \\
\phantom{} [1,1] & (D)
\end{array}
\]
States of types (B) and (C) belong to a single orbit.
The pair stays intact and acts as a permanent quasiparticle, and
the lattice grows exactly as in the 1 vs. 2 case:

\[
\begin{array}{rlc}
(C) & [1;{\bf g_{12}=1},{\bf g_{23}},g_{31};0] \\
  & & \downarrow {\bf +1} \\
  & [1;g_{31},{\bf g_{12}=1}, g_{23}+1;1] \\
  & & \downarrow -1 \\
(B) & [1;{\bf g_{12}=1}, g_{23}+1,{\bf g_{31}-1};0] \\
  & & \downarrow {\bf +1} \\
  & [0; {\bf g_{31}-1}, {\bf g_{12}=1},{\bf g_{23}+2};1]\\
  & & \downarrow +1 \\
(C) & [1; {\bf g_{12}=1},{\bf g_{23}+2},g_{31};0]
\end{array}
\]

The pattern of type (A) is given below. 
Note that after eight interactions, we have a state 
of type (D), so type (D) is contained in that orbit.
As long as no other pairs appear,
type (A) forms a parity-periodic orbit of length sixteen.
$\Delta L=-2$ after one such orbit:
the gap between particles $l_2$, $l_3$ decreases by two.
Pairs can appear in state (d) if $g_{23}+1=1$, 
in state (j) if $g_{31}+1=1$ and in state (o) if $g_{23}-1=1$.

Suppose $g_{31}+1=1$ in (j). This is a pair as in type
(C), so from here on, the orbit is growing.

Let $g_{23}-2k+1=1$. The change in the pattern occurs
in the $k$-th run through the orbit (a)-(p).
For $k=0$ state (d) contains a pair of type (C).
For $k>0$ state (o) contains a pair of type (D),
$[1;1,1,1]$ with first {\bf and} third gap length equal to one.
This state evolves to state (j) with first gap of length
one: a type (C) pair has formed. 
We already know that type (C) belongs to a growing orbit.

So systems of type (A) and (D) produce a
parity-periodic orbit of decreasing length. 
As soon as a second pair forms, we observe a transition via (D)
to the growing orbit of type (B).

Type (A):
\[
\begin{array}{clc}
(a)& [1;{\bf g_{12}=1},g_{23},g_{31};1] \\
  & & \downarrow {\bf +1} \\
(b)& [0; g_{31},{\bf g_{12}=1},{\bf g_{23}+1};0]\\
  & & \downarrow +1 \\
(c)& [0;{\bf g_{12}=1},{\bf g_{23}+1},{\bf g_{31}+1};1] \\
  & & \downarrow +1 \\
(d)& [1;{\bf g_{23}+1},{\bf g_{31}+1}, g_{12}+1;0] \\
  & & \downarrow -1 \\
(e)& [0;{\bf g_{31}+1}, g_{12}+1, g_{23};1] \\
  & & \downarrow +1 \\
(f)& [1;g_{12}+1,g_{23},g_{31}+2;0] \\
  & & \downarrow -1 \\
(g)& [1;g_{23}, g_{31}+2,{\bf g_{12}=1};1]\\
  & & \downarrow -1 \\
(h)& [1; g_{31}+2,{\bf g_{12}=1},{\bf g_{23}-1};0] \\
  & & \downarrow -1 \\
(i)& [1;{\bf g_{12}=1},{\bf g_{23}-1},{\bf g_{31}+1};1]\\
  & & \downarrow {\bf +1} \\
(j)& [1; {\bf g_{31}+1},{\bf g_{12}=1}, g_{23};0] \\
  & & \downarrow -1 \\
(k)& [0;{\bf g_{12}=1},g_{23},g_{31};1] \\
  & & \downarrow +1 \\
(l)& [1;g_{23},g_{31},g_{12}+1;0]\\
  & & \downarrow -1 \\
(m)& [1;g_{31},g_{12}+1,{\bf g_{23}-1};1]\\
  & & \downarrow -1 \\
(n)& [1;g_{12}+1,{\bf g_{23}-1},{\bf g_{31}-1};0]\\
  & & \downarrow -1 \\
(o)& [1;{\bf g_{23}-1},{\bf g_{31}-1},{\bf g_{12}=1};1]\\
  & & \downarrow -1\\
(p)& [0;{\bf g_{31}-1},{\bf g_{12}=1},g_{23}-2;0]\\
  & & \downarrow +1 \\
(a')& [1;{\bf g_{12}=1},g_{23}-2,g_{31};1]
\end{array}
\]

\end{ex}

%
%
\subsection{Two Against Two}
%
%

This case needs more information: now there are two right-movers and
they interact in a row.
We have to adapt the notation and keep track of the positions of the
right-movers.

\begin{figure}
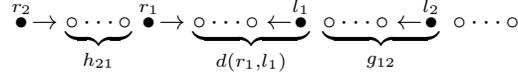
\[
\quad\stackrel{r_2}{\bullet}\to
\underbrace{\circ
\dots\circ}_{h_{21}}\stackrel{r_1}{\bullet}\to
\underbrace{\circ\dots\circ
\leftarrow \stackrel{l_1}{\bullet}}_{d(r_1,l_1)}
\ \underbrace{\circ\dots\circ\leftarrow
\stackrel{l_2}{\bullet}}_{g_{12}}\
\circ\cdots\circ
\]
\caption{Notation for two right-movers vs. two left-movers.}
\end{figure}

Suppose $r_1$ interacts first with $l_1$. Then there
are several possibilities:

\begin{itemize}
\item[(i)]
Particle
$r_1$ interacts next with $l_2$, then
$r_2$ interacts with $l_1$.
\item[(ii)]
There is an interlaced pattern: particle $r_2$
interacts with $l_1$, then $r_1$ interacts
with $l_2$, then $r_2$ with $l_2$.
\item[(iii)]
Particles $r_1$ and $l_2$ interact at the same
time step as $r_2$ with $l_1$.
\end{itemize}

It can be checked that it does not matter which interaction
is first (as long as no pairs occur). 
This is a reflection of the locality of the scattering.

We start by looking at the cases where no pairs appear. 
In order to describe the evolution of the system, we give pairs of $4$-tuples:
$
[d_{11};g_{12},g_{21};|L|]\quad
[d_{21};g_{12},g_{21};|L|].
$

In contrast to the previous notation, we give the actual separation
between the right-mover $r_i$ and the next left-mover $l_j$ that $r_i$ faces. 
Let $h_{21}:=d(r_2,r_1)$ be the gap between the two right-movers.
As before, bold font is used to denote odd length.
Furthermore, we add an arrow $\pm 1\downarrow$ to show
the change of the length of the lattice.
The arrows also indicate which pair is interacting.

Figure 4 shows a ``noninterlaced" state in which there is no left-mover 
between the two right-movers, and vice-versa.
We can always assume such an initial state, unless the four particles are
located symmetrically around the ring: right-movers diametrically opposite
one another, left-movers likewise along a perpendicular diameter.
It is easy to see that these exceptional initial states evolve as periodic
orbits, and we will not consider them further.

Note that $d_{21}=d_{11}+h_{21}$.
We know that the order of the interactions does not matter. 
Thus in the example below we can fix the order of interactions by assuming 
$h_{21}\le g_{12}-2$ in (2) 
and $h_{21}+1\le g_{21}-2$ in (4). 

We first give the two periodic orbits.
Then we will describe the growing orbits and those with decreasing length.
In the end we discuss the cases with pairs.

\begin{ex}\label{ex:2-2constI}

We start with a system where $g_{12}$ is even, $g_{21}$ odd 
(i.e. the lattice has odd length). 
Also let $h_{21}$ and $d_{11}$ be even.
Recall that notation such as $g_{12}$ below denotes this gap in the 
{\it initial} state (1); as the gaps change during evolution their 
current values are indicated by their positions.

\[
\begin{array}{clclc}
  & \mbox{Particle $r_1$} & & \mbox{Particle $r_2$} \\
& \\
(1)&[d_{11};g_{12},{\bf g_{21}};1]
& & [h_{21}+d_{11};g_{12},{\bf g_{21}};1] \\
  & & \downarrow +1 & \\
(2)&[g_{12};{\bf g_{21}},{\bf g_{12}+1};0]
& & [h_{21};{\bf g_{12}+1},{\bf g_{21}};0]\\
  & & & & \downarrow +1 \\
(3)&[g_{12}-h_{21}; {\bf g_{21}},g_{12}+2;1]
& & [{\bf g_{12}+1};g_{12}+2,{\bf g_{21}};1] \\
  & & \downarrow +1 & \\
(4)&[{\bf g_{21}};g_{12}+2,g_{21}+1;0]
& & [{\bf h_{21}+1};g_{21}+1,g_{12}+2;0]\\
  & & & & \downarrow -1 \\
(5)&[{\bf g_{21}-h_{21}-2};g_{12}+2,{\bf g_{21}};1]
& & [{\bf g_{21}};g_{12}+2,{\bf g_{21}};1]\\
  & & \downarrow -1 & \\
(6)&[{\bf g_{12}+1};{\bf g_{21}},{\bf g_{12}+1};0]
& & [{\bf h_{21}+1};{\bf g_{12}+1},{\bf g_{21}};0]\\
  & & & & \downarrow -1 \\
(7)&[{\bf g_{12}-h_{21}-1};{\bf g_{21}},g_{12};1]
& & [g_{12};{\bf g_{21}},g_{12};1]\\
  & & \downarrow -1 & \\
(8)&[g_{21}-1;g_{12},g_{21}-1;0]
& & [h_{21};g_{21}-1,g_{12};0]\\
  & & & & \downarrow +1 \\
(1)'&[g_{21}-h_{21}-1;g_{12},{\bf g_{21}};1]
& & [g_{21}-1;g_{12},{\bf g_{21}};1]
\end{array}
\]

The assumption $g_{21}\ge 3$ (i.e. the gap between
$l_2$ and $l_1$ is at least three) 
ensures that the exclusion rule does not apply in step (7). 
\end{ex}

There are 16 possible parity combinations for $h_{21},d_{11},g_{12},g_{21}$.
In that language, the parities that occur 
in Example~\ref{ex:2-2constI} above are 

$[0,0,0,1]$ in (1), $[1,0,1,0]$ in (3),
$[0,1,0,1]$ in (5) and $[1,1,1,0]$ in (7).

\begin{ex}
Next we assume $d_{11}$ to be even,
$h_{21}$, $g_{21}$ even and $g_{12}$ odd.
\[
\begin{array}{clclc}
  & \mbox{Particle $r_1$} & & \mbox{Particle $r_2$} \\
& \\
(1)&[d_{11};{\bf g_{12}},g_{21};1]
& & [h_{21}+d_{11};{\bf g_{12}},g_{21};1] \\
  & & \downarrow +1 & \\
(2)&[{\bf g_{12}};g_{21},g_{12}+1;0]
& & [h_{21};g_{12}+1,g_{21};0] \\
  & & & & \downarrow +1 \\
(3)&[{\bf g_{12}-h_{21}};g_{21},{\bf g_{12}+2};1]
& & [g_{12}+1;g_{21},{\bf g_{12}+2};1]\\
  & & \downarrow -1 \\
(4)&[g_{21}-1;{\bf g_{12}+2},{\bf g_{21}-1};0]
& & [h_{21};{\bf g_{21}-1},{\bf g_{12}+2};0]\\
  & & & &\downarrow +1 \\
(5)&[{\bf g_{21}-h_{21}-1};{\bf g_{12}+2},g_{21};1]
& & [{\bf g_{21}-1};{\bf g_{12}+2},g_{21};1] \\
  & & \downarrow -1 \\
(6)&[g_{12}+1;g_{21},g_{12}+1;0]
& & [{\bf h_{21}-1};g_{12}+1,g_{21};0] \\
  & & & & \downarrow -1\\
(7)&[g_{12}-h_{21}+1;g_{21},{\bf g_{12}};1]
& & [{\bf g_{12}};g_{21},{\bf g_{12}};1] \\
  & & \downarrow +1 \\
(8)&[g_{21};{\bf g_{12}},{\bf g_{21}+1};0]
& & [{\bf h_{21}-1};{\bf g_{21}+1},{\bf g_{12}};0] \\
  & & & & \downarrow -1 \\
(1)'&[g_{21}-h_{21};{\bf g_{12}},g_{21};1]
& & [g_{21};{\bf g_{12}},g_{21};1]
\end{array}
\]

This is again a periodic orbit.
The only situations where the exclusion rule can apply are in state
(5) if $g_{12}+2=1$ (i.e. $d(l_1,l_2)=1$)
or in state (3) if $g_{12}+1-(g_{12}-h_{21})=h_{21}+1=1$
(i.e. $d(r_2,r_1)=1$).
Note that the odd numbered states are the interlaced
ones.
The parity configurations of the distances
$d(r_{i+1},r_i), d(r_i,l_j), d(l_j,l_{j+1})$,
$d(l_{j+1},l_{j+2})$ are

$[0,0,1,0]$ in (1), $[1,1,0,1]$ in (3),
$[0,1,1,0]$ in (5) and $[1,1,0,1]$ in (7).
\end{ex}

Now we describe the systems where the length of
the lattice is increasing.
\begin{ex}
Let $h_{21}$ and $d_{11}$ be even,
$g_{12}$ and $g_{21}$ odd.
\[
\begin{array}{clclc}
  & \mbox{Particle $r_1$} & & \mbox{Particle $r_2$} \\
& \\
(1)&[d_{11};{\bf g_{12}},{\bf g_{21}};0]
& & [h_{21}+d_{11};{\bf g_{12}},{\bf g_{21}};0] \\
  & & \downarrow +1 & \\
(2)&[{\bf g_{12}};{\bf g_{21}},g_{12}+1;1]
& & [h_{21};g_{12}+1,{\bf g_{21}};1] \\
  & & & & \downarrow +1 \\
(3)&[{\bf g_{12}-h_{21}};{\bf g_{21}},{\bf g_{12}+2};0]
& & [g_{12}+1;{\bf g_{21}},{\bf g_{12}+2};0] \\
  & & \downarrow -1\\
(4)&[g_{21}-1;{\bf g_{12}+2},g_{21}-1;1]
& & [h_{21};g_{21}-1,{\bf g_{12}+2};1] \\
  & & & & \downarrow +1 \\
(1)'&[g_{21}-h_{21}-1;{\bf g_{12}+2},{\bf g_{21}};0]
& & [g_{21}-1;{\bf g_{12}+2},{\bf g_{21}};0]
\end{array}
\]

After four interactions we return to the initial parity configuration,
with $\Delta L=2$ (the gap $d(l_1,l_2)$ grows by two).
In the noninterlaced states we have the following
parity configurations of the distances
$d(r_{i+1},r_i), d(r_i,l_j), d(l_j,l_{j+1})$,
$d(l_{j+1},l_{j+2})$:

$[0,0,1,1]$ in (1) and $[1,1,1,1]$ in (3).
\end{ex}

\begin{ex}\label{ex:growingI}
In this case, $h_{21}$ is odd, $d_{11}$ even,
$g_{12}$ and $g_{21}$ even.

\[
\begin{array}{clclc}
  & \mbox{Particle $r_1$} & & \mbox{Particle $r_2$} \\
& \\
(1)&[d_{11};g_{12}, g_{21};0]
& & [{\bf h_{21}+d_{11}};g_{12}, g_{21};0] \\
  & & \downarrow +1 & \\
(2)&[g_{12};g_{21},{\bf g_{12}+1};1]
& & [{\bf h_{21}};{\bf g_{12}+1},g_{21};1]\\
  & & & & \downarrow -1 \\
(3)&[g_{12}-h_{21}-1;g_{21},g_{12};0]
& & [g_{12};g_{21},g_{12};0] \\
  & & \downarrow +1 \\
(4)&[g_{21};g_{12},{\bf g_{21}+1};1]
& & [h_{21}+1;{\bf g_{21}+1},g_{12};1]\\
  & & & & \downarrow +1\\
(1)'&[g_{21}-h_{21}-1;g_{12}, g_{21}+2;0]
& & [{\bf g_{21}+1};g_{12},g_{21}+2;0]
\end{array}
\]

This is another growing lattice. 
After four interactions, the gap between $l_2$ and $l_1$ has grown by two,
with the other gaps unchanged.
The parities are $[1,0,0,0]$ in (1) and $[0,0,0,0]$ in (3).
\end{ex}

There are four remaining parity combinations. 
They belong to two systems with decreasing length. 
We describe them in the two examples below:
\begin{ex}
Let $h_{21}$, $g_{12}$ and $g_{21}$ be even,
$d_{11}$ odd.
\[
\begin{array}{clclc}
  & \mbox{Particle $r_1$} & & \mbox{Particle $r_2$} \\
& \\
(1)&[{\bf d_{11}}; g_{12}, g_{21};0]
& & [{\bf h_{21}+d_{11}};g_{12}, g_{21};0] \\
  & & \downarrow -1 & \\
(2)&[{\bf g_{12}-1};g_{21},{\bf g_{12}-1};1]
& & [{\bf h_{21}-1};{\bf g_{12}-1},g_{21};1] \\
  & & & & \downarrow -1 \\
(3)&[{\bf g_{21}-h_{21}-1};g_{21},g_{12}-2;0]
& & [g_{12}-2;g_{21},g_{12}-2;0] \\
  & & \downarrow -1 \\
(4)&[{\bf g_{21}-1};g_{12}-2,{\bf g_{21}-1};1]
& & [h_{21}-2;{\bf g_{21}-1},g_{12}-2;1] \\
  & & & & \downarrow +1 \\
(1)'&[{\bf g_{21}-h_{21}+1}; g_{12}-2,g_{21};0]
& & [{\bf g_{21}-1};g_{12}-2,g_{21};0]
\end{array}
\]
After four interactions we return to the same
parities but with $\Delta L=-2$ (gap $g_{12}\mapsto g_{12}-2$). 
In other words, this system is shrinking as long as
no pairs occur. 
A pair can form in state (3). 
There the distance between particles $r_2$ and $r_1$
after $k$ cycles (1)-(4) is $g_{12}-2k-(g_{21}-h_{21}-1)$, 
which will eventually shrink to $1$.
In this case we obtain pairs as in the system with label b) of
Subsection~\ref{sse:neighbors-2-2} by
switching the roles of the left-movers and the right-movers.
The parities of this example are
$[0,1,0,0]$ in (1) and
$[1,1,0,0]$ in (3).
\end{ex}
\begin{ex}
Finally, let $d_{11}$, $g_{12}$, $g_{21}$ be odd and
$h_{21}$ be even.
\[
\begin{array}{clclc}
  & \mbox{Particle $r_1$} & & \mbox{Particle $r_2$} \\
& \\
(1)&[{\bf d_{11}};{\bf g_{12}},{\bf g_{21}};0]
& & [{\bf h_{21}+d_{11}};{\bf g_{12}},{\bf g_{21}};0] \\
  & & \downarrow -1 & \\
(2)&[g_{12}-1;{\bf g_{21}},g_{12}-1;1]
& & [{\bf h_{21}-1};g_{12}-1,{\bf g_{21}};1]\\
  & & & & \downarrow -1 \\
(3)&[g_{12}-h_{21}-1;{\bf g_{21}},{\bf g_{12}-2};0]
& & [{\bf g_{12}};{\bf g_{21}},g_{12}-2;0]\\
  & & \downarrow +1 \\
(4)&[{\bf g_{21}};{\bf g_{12}-2},g_{21}+1;1]
& & [{\bf h_{21}-1};g_{21}+1,{\bf g_{12}-2};1]\\
  & & & & \downarrow -1\\
(1)'&[{\bf g_{21}-h_{21}};{\bf g_{12}-2},{\bf g_{21}};0]
& & [{\bf g_{21}};{\bf g_{12}-2},{\bf g_{21}};0]
\end{array}
\]
After four interactions, the length of the lattice
has decreased by two (gap $g_{12}\mapsto g_{12}-2$).
This is a shrinking system as long as no pairs
appear. 
A pair appears in state (1), as soon as $g_{12}-2k=1$ 
[i.e. after $k$ cycles (1)-(4)],
when we obtain a pair as in the system with label b),
cf. Subsection~\ref{sse:neighbors-2-2}.
The parities of this example are $[0,1,1,1]$ in (1)
and $[1,0,1,1]$ in (3).
\end{ex}

We now describe the patterns when pairs are present initially or form
during the evolution.
%
%
\subsection{Two Against Two With Pairs}
\label{sse:neighbors-2-2}
%

\begin{figure}
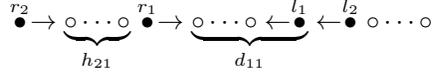

\[
\quad\stackrel{r_2}{\bullet}\to
\underbrace{\circ
\dots\circ}_{h_{21}}\stackrel{r_1}{\bullet}\to
\underbrace{\circ\dots\circ
\leftarrow \stackrel{l_1}{\bullet}}_{d_{11}}
\leftarrow\stackrel{l_2}{\bullet}
\circ\cdots\circ
\]
\caption{Notation for two vs. two with pairs.}
\end{figure}

There are six different configurations with pairs.
The gap $d_{11}$ has to be odd if the pair is to affect the evolution
of the system, and we assume $g_{12}=1$.
Then we have the following parity combinations for the remaining distances:

$$\begin{array}{cclc}
\mbox{Label} & h_{21} & & g_{21} \\
\hline
a) & 0 & & 0 \\
b) & 0 & & 1 \\
c) & 1 & \mbox{with } h_{21}>1 & 0 \\
d) & 1 & \mbox{with } h_{21}>1 & 1 \\
e) & 1 & \mbox{with } h_{21}=1 & 0 \\
f) & 1 & \mbox{with } h_{21}=1 & 1
\end{array}$$

\begin{ex}[Cases a),d) and c)]
We start with $h_{21}$ and $g_{21}$ even (i.e. a lattice of odd length). 
Assume that no other pair forms.
The corresponding conditions are
$g_{21}-(h_{21}-1)\ge 3$ in state (2),
and $g_{21}-(g_{21}-h_{21}-1)=h_{21}+1\ge 3$
in state (7).
We will discuss these other cases below.
\[
\begin{array}{clclc}
  & \mbox{Particle $r_1$} & & \mbox{Particle $r_2$} \\
& \\
(1)&[{\bf d_{11}};{\bf g_{12}=1},g_{21};1]
& & [{\bf h_{21}+d_{11}};{\bf g_{12}=1},g_{21};1] \\
  & & \downarrow {\bf +1}  \\
(2)&[g_{21};{\bf g_{12}=1},{\bf g_{21}+1};0]
& & [{\bf h_{21}-1};{\bf g_{12}=1},{\bf g_{21}+1};0]\\
  & & & & \downarrow {\bf +1} \\
(3)&[g_{21}-h_{21}; {\bf g_{12}=1},g_{21}+2;1]
& & [{\bf g_{21}+1};{\bf g_{12}=1},g_{21}+2;1]\\
  & & \downarrow +1 \\
(4)&[{\bf g_{12}=1};g_{21}+2,g_{12}+1;0]
& & [{\bf h_{21}+1};g_{12}+1,g_{21}+2;0]\\
  & & \downarrow -1 \\
(5)&[{\bf g_{21}+1};g_{12}+1,{\bf g_{21}+1};1]
& & [{\bf h_{21}-1};g_{12}+1,{\bf g_{21}+1};1]\\
  & & & & \downarrow -1 \\
(6)&[{\bf g_{21}-h_{21}+1};{\bf g_{12}=1},{\bf g_{21}+1};0]
& & [{\bf g_{12}=1};{\bf g_{21}+1},{\bf g_{12}=1};0]\\
  & & & & \downarrow -1\\
(7)&[{\bf g_{21}-h_{21}-1};{\bf g_{12}=1},g_{21};1]
& & [g_{21};{\bf g_{12}=1},g_{21};1]\\
  & & \downarrow {\bf +1}\\
(8)&[g_{21}; {\bf g_{12}=1},{\bf g_{21}+1};0]
& & [h_{21};{\bf g_{12}=1},{\bf g_{21}+1};0]\\
  & & & & \downarrow +1\\
(9)&[g_{21}-h_{21};g_{12}+1,{\bf g_{21}+1};1]
& & [{\bf g_{12}=1};{\bf g_{21}+1},g_{12}+1;1]\\
  & & & & \downarrow -1\\
(10)&[g_{21}-h_{21}-g_{12}-1;g_{12}+1,g_{21};0]
& & [g_{21};g_{12}+1,g_{21};0]\\
\end{array}
\]

The first observation is that state (2) is
as case d) (with particles $r_1$, $r_2$
switched) by the assumption that the gap
between $r_1$ and $r_2$ is at least $3$.

Similarly, state (7) is as case c)
The distance between $r_2$ and $r_1$ is odd
and at least $3$ (by the assumptions),
the distance between $l_2$ and $l_1$ is even.

Finally note that state (10) is as state (3) in
Example~\ref{ex:growingI}. 
This means that the system evolves into a growing lattice as in
Example~\ref{ex:growingI}.

Consider now other pairs of nearest neighbors.

Let $g_{21}=h_{21}$ (i.e.
$g_{21}-(h_{21}-1)=1$ in (2)). 
Then the state (2) has the same parities as case f): 
the right-movers form a pair, as do the left-movers, and the gap between 
the particles $r_2$ and $l_1$ is odd. 
So for $g_{21}=h_{21}$ the system will evolve as f).

Let $h_{21}+1=1$. 
Then state (7) has the same parities as case e).
\end{ex}

\begin{ex}[Case b)]
We start with $h_{21}$ even and $g_{21}$ odd (i.e.
a lattice of even length).
\[
\begin{array}{clclc}
  & \mbox{Particle $r_1$} & & \mbox{Particle $r_2$} \\
& \\
(1)&[{\bf d_{11}};{\bf g_{12}=1},{\bf g_{21}};0]
& & [{\bf h_{21}+d_{11}};{\bf g_{12}=1},{\bf g_{21}};0] \\
  & & \downarrow {\bf +1}\\
(2)&[{\bf g_{21}};{\bf g_{21}=1},g_{21}+1;1]
& & [{\bf h_{21}-1};{\bf g_{12}=1},g_{21}+1;1]\\
  & & & & \downarrow {\bf +1}
\end{array}
\]
Note that state (2) is as in a).
\end{ex}
It remains to discuss cases e) and f). 
These are the cases where two pairs of nearest neighbors
face each other with an odd distance between them. 
That means we have $g_{12}=h_{21}=1$, and necessarily $g_{21}\ge 3$.

\begin{ex}\label{ex:two-pairs-oddL}
Let particles $r_2$, $r_1$ be paired as well as particles $l_1$, $l_2$. 
Assume that $g_{21}$ is even (i.e. lattice of odd length).
So $g_{21}$ is at least equal to four.
\[
\begin{array}{clclc}
  & \mbox{Particle $r_1$} & & \mbox{Particle $r_2$} \\
& \\
(1)&[{\bf d_{11}};{\bf g_{12}=1},g_{21};1]
& & [h_{21}+d_{11};{\bf g_{12}=1},g_{21};1] \\
  & & \downarrow {\bf +1} & & \downarrow {\bf +1}\\
(2)&[g_{21};g_{12}+1,{\bf g_{21}+1};1]
& & [{\bf h_{21}=1};{\bf g_{21}+1},g_{12}+1;1] \\
  & & & & \downarrow -1\\
(3)&[g_{21}-h_{21}-1;g_{12}+1,g_{21};0]
& & [g_{21};g_{12}+1,g_{21};0]\\
\end{array}
\]
Note that state (3) has the same parities as
state (3) in Example~\ref{ex:growingI}. 
So the system evolves into a growing orbit as in Example~\ref{ex:growingI}.
\end{ex}

\begin{ex}\label{ex:two-pairs-evenL}
Let $r_2$, $r_1$  and $l_1$, $l_2$ each be pairs, 
with $g_{21}$ odd ($g_{21}\ge 3$).
The length of the lattice is then even.
\[
\begin{array}{clclc}
  & \mbox{Particle $r_1$} & & \mbox{Particle $r_2$} \\
& \\
(1)&[{\bf d_{11}};{\bf g_{12}=1},{\bf g_{21}};0]
& & [h_{21}+d_{11};{\bf g_{12}=1},{\bf g_{21}};0] \\
  & & \downarrow {\bf +1} & & \downarrow {\bf +1}\\
(2)&[{\bf g_{21}};g_{12}+1, g_{21}+1;0]
& & [{\bf h_{21}=1}; g_{21}+1,g_{12}+1;0] \\
  & & & & \downarrow -1\\
(3)&[{\bf g_{21}-h_{21}-1};g_{12}+1,{\bf g_{21}};1]
& & [{\bf g_{21}};g_{12}+1,{\bf g_{21}};1]\\
  & & \downarrow -1 \\
(4)&[{\bf g_{12}=1};{\bf g_{21}},{\bf g_{12}=1};0]
& & [{\bf h_{21}};{\bf g_{12}=1},{\bf g_{21}};0]\\
  & & & & \downarrow {\bf +1} \\
(5)&[g_{21}-h_{21}; g_{21}+1,{\bf g_{12}},1]
& & [{\bf g_{21}};{\bf g_{12}=1},g_{21}+1;1]\\
  & & \downarrow +1 \\
(6)&[{\bf g_{12}=1};g_{21}+1,g_{12}+1;0]
& & [{\bf h_{21}};g_{12}+1,g_{21}+1;0]\\
  & & \downarrow -1 & & \downarrow -1 \\
(7)&[{\bf g_{21}};{\bf g_{12}=1},{\bf g_{21}};0]
& & [{\bf g_{12}=1};{\bf g_{21}},{\bf g_{12}=1};0] \\
  & & & & \downarrow -1 \\
(8)&[{\bf g_{21}-g_{12}-1};{\bf g_{12}=1},g_{21}-1;1]
& & [g_{21}-1;{\bf g_{12}=1},g_{21}-1;1]
\end{array}
\]
Note that state (8) has the same parities as
state (1) in Example~\ref{ex:two-pairs-oddL}.
So the system will switch to that example and
then to a growing orbit (as in Example~\ref{ex:growingI}).
\end{ex}


\section{Conclusions}


In this paper we have studied the lattice gas model with
dynamical geometry introduced by Hasslacher and Meyer \cite{hm98}.
We first gave a general discussion of the pair-free evolution, establishing
the importance of parity and the existence of many periodic orbits.
The $\sqrt{t}$ growth of the lattice length observed in simulations is
expected whenever the length grows by a constant amount, on average, as
the particles complete one circuit.
This is the case without pairs, and it is plausible that it is the general
asymptotic behavior, but we have no proof of this.
At present we can include the effects of pairs only by an exhaustive
case-by-case analysis, which we carried out for systems of at most four
particles.
The pairs can form permanent ``bound states", and it seems promising to 
view them as quasiparticles.
With an effective description of these quasiparticles, it should be possible
to solve the dynamics of this model completely.

Although the microscopic dynamics of this model is reversible, it exhibits
macroscopic ``irreversibility" as the length grows for ``most" states.
It might be interesting to quantify what fraction of initial states having
given $N_L, N_R, L$ ultimately grow.

The natural conjecture regarding the asymptotics of the model is that all
nonperiodic solutions grow as $\sqrt{t}$, reflecting a constant average
growth per circuit.
Is this really true?
Perhaps there are solutions with a characteristic time scale much longer than 
one circuit.
For example, imagine a solution in which the length initially 
has average growth zero per circuit.
A randomly chosen right-right gap would have size of order 
$L/N_R$ and could shrink to form a pair in a time $\sim L^2/N_RN_L$.
If this pair alters the evolution to produce constant average growth 
on this time scale,
then $dL/dt \sim 1/L^2$ leads to $L \sim t^{1/3}$.
Are there solutions with this behavior?
Do the fluctuations in a solution growing as $\sqrt{t}$ include intervals
when the growth is as $t^{1/3}$, correlated with the formation and
destruction of pairs?

Finally, it should be straightforward to quantize this model lattice gas, 
generalizing \cite{m96,m97}.
The Hilbert space would be the direct sum 
$H = \oplus_{L=1}^\infty H_L$ of the Hilbert spaces for lattices of all
fixed lengths $L$.
Advection occurs within each $H_L$, but scattering causes transitions
between them.
The quantized model would be similar to that of \cite{bcv05}, but with
evolution in discrete rather than continuous time.
It is possible that the model, at least with few particles, is solvable
by Bethe ansatz or other methods.


\end{document}